\documentclass[12pt]{article}

\usepackage[utf8]{inputenc}
\usepackage{geometry}
\usepackage{graphicx}
\usepackage{array}
\usepackage{slashed}
\usepackage{amsmath}
\usepackage{amsfonts}
\usepackage{amssymb}
\usepackage{physics}
\usepackage{mathtools}
\usepackage{placeins}
\usepackage[usenames, dvipsnames]{color}
\usepackage{caption}
\usepackage{subcaption}
\usepackage{dsfont}
\usepackage[cm]{fullpage}
\usepackage{cite}
\usepackage{epstopdf}
\usepackage{comment}
\usepackage{hyperref}
\usepackage{enumitem}
\usepackage{cancel}
\usepackage{mathrsfs}

\usepackage{cleveref}
\crefname{section}{§}{§§}
\Crefname{section}{§}{§§}
\usepackage{booktabs} 

\numberwithin{equation}{section}

\def\p{\partial}

\def\0{{(0)}}
\def\1{{(1)}}
\def\2{{(2)}}

\def\<{\langle }
\def\>{\rangle }

\newcommand{\bea}{\begin{eqnarray}}

\newcommand{\eea}{\end{eqnarray}}

\newcommand{\be}{\begin{equation}}
\newcommand{\ee}{\end{equation}}
\newcommand{\ba}{\begin{align}}
\newcommand{\ea}{\end{align}}

   \makeatletter
  \let\over=\@@over \let\overwithdelims=\@@overwithdelims
  \let\atop=\@@atop \let\atopwithdelims=\@@atopwithdelims
  \let\above=\@@above \let\abovewithdelims=\@@abovewithdelims
\renewcommand\section{\@startsection {section}{1}{\z@}%
                                   {-3.5ex \@plus -1ex \@minus -.2ex}
                                   {2.3ex \@plus.2ex}%
                                   {\normalfont\large\bfseries}}

\renewcommand\subsection{\@startsection{subsection}{2}{\z@}%
                                     {-3.25ex\@plus -1ex \@minus -.2ex}%
                                     {1.5ex \@plus .2ex}%
                                     {\normalfont\bfseries}}

\newcommand{\beq}{\begin{equation}}
\newcommand{\eeq}{\end{equation}}
\newcommand{\beqa}{\begin{eqnarray}}
\newcommand{\eeqa}{\end{eqnarray}}
\newcommand{\beqar}{\begin{eqnarray*}}

\def\[{\[}
\def\]{\]}

\newcommand{\bd}[1]{\begin{fmffile}{#1}\begin{fmfgraph*}}
\newcommand{\ed}{\end{fmfgraph*}\end{fmffile}}

\begin{document}

\begin{titlepage}

\unitlength = 1mm~\\
\vskip 1cm
\begin{center}

{\LARGE{\textsc{Three-Generation Super No-Scale Models \\[0.3cm] in Heterotic Superstrings}}}

\vspace{0.8cm}
Ioannis Florakis\,{}\footnote{{\tt iflorakis@uoi.gr}}, John Rizos\,{}\footnote{\tt irizos@uoi.gr} and Konstantinos Violaris-Gountonis\,{}\footnote{\tt k.violaris@uoi.gr}

\vspace{1cm}

{\it  Department of Physics, University of Ioannina\\ GR45110 Ioannina, Greece 

}

\vspace{0.8cm}

\begin{abstract}
We derive the conditions for the one-loop contribution to the cosmological constant to be exponentially suppressed in a 
class of heterotic string compactifications with three generations of chiral matter, 
where supersymmetry is spontaneously broken \`a la Scherk-Schwarz. Using techniques of partial unfolding based on the Hecke congruence subgroup $\Gamma_0(2)$ of the modular group to extract the leading asymptotics, we show that the super no-scale condition $n_B=n_F$ between the degeneracies of massless bosons $n_B$ and fermions $n_F$ in the full string spectrum is necessary but not sufficient, and needs to be supplemented by additional conditions which we identify.
We use these results to construct three-generation Pati--Salam models with interesting 
phenomenological characteristics.

\end{abstract}

\setcounter{footnote}{0}

\vspace{1.0cm}

\end{center}

\end{titlepage}

\pagestyle{empty}
\pagestyle{plain}

\def\vx{{\vec x}}
\def\p{\partial}
\def\po{$\cal P_O$}

\pagenumbering{arabic}



\section{Introduction}

Supersymmetry may be spontaneously broken within a fully-fledged string theory setup admitting an exact worldsheet 
description via coordinate-dependent compactifications \cite{Rohm:1983aq,Kounnas:1988ye,Ferrara:1988jx,Kounnas:1989dk}, which essentially realise the stringy analogue of the 
Scherk-Schwarz mechanism \cite{Scherk:1978ta,Scherk:1979zr}. This involves coupling a boost of the fermionic charge lattice of the theory with an 
accompanying shift along a 1-cycle of the compactification space. Alternatively, it may be reformulated as a 
freely-acting $\mathbb Z_2$ orbifold, involving the spacetime fermion parity $(-1)^F$, together with an order-two shift 
along a compact dimension, the characteristic size $R$ of which controls the breaking scale $m_{3/2}\sim 1/R$. 

The resulting string spectrum contains scalars with moduli-dependent masses, which may potentially become tachyonic 
when the Scherk-Schwarz radius is close to the string scale $R\sim \sqrt{\alpha'}$, signalling a tree-level 
destabilisation of the theory \cite{Atick:1988si,Kutasov:1990sv,Dienes:1994np,Angelantonj:2008fz,Angelantonj:2010ic,Abel:2016hgy}. Working in regions $R\gg \sqrt{\alpha'}$ of moduli space sufficiently away from the 
string scale typically secures the absence of tachyonic modes, although a suitable stabilisation mechanism is still 
required. A further discrepancy arises from the fact that the cosmological constant receives sizeable contributions in 
contradiction with its observed value. This is because the effective potential $V_{\rm eff}$ of the theory is no longer 
super-protected and receives quantum corrections at all loop orders. Hence, although the tree-level theory is a 
no-scale model with a vanishing minimum of the potential \cite{Cremmer:1983bf}, radiative corrections spoil this property already at one 
loop and induce a non-trivial back-reaction. Setting $\alpha'=1$ and at sufficiently large radius, the one-loop contribution takes the generic form \cite{Antoniadis:1990ew},
\begin{equation}
	V_{\rm eff} \simeq -\frac{N}{R^4} + \mathcal{O}(e^{-\lambda R})\,,
\label{Vgeneric}
\end{equation}
where $N$ is a constant proportional to the degeneracies of the massless string spectrum and $\lambda$ a positive 
constant of order one. It is then easy to see that, even if we set the compactification radius $R$ to the TeV range and $N\sim 
1$, the observed value of the cosmological constant is overshot by more than 30 orders of magnitude. 

Actually, exceptions to this largely universal behaviour exist. On the one hand, there are examples of non-supersymmetric theories where the one-loop vacuum energy vanishes \cite{Kachru:1998hd,Angelantonj:1999gm,Harvey:1998rc,Shiu:1998he,Angelantonj:2004cm}. On the other hand, there are theories where only the coefficient of the power law term in \eqref{Vgeneric} vanishes identically, in which case the one-loop contribution to the effective potential is 
exponentially suppressed. Such theories where the no-scale properties are effectively extended to the one-loop 
level were termed ``super no-scale" models in recent works \cite{Kounnas:2016gmz,Florakis:2016ani,Kounnas:2017mad,Florakis:2021bws}. The corresponding condition $N=0$ was further interpreted 
as the requirement $n_B-n_F=0$ of having an equal number of bosonic and fermionic states in the massless spectrum of 
the theory.

In previous work \cite{Florakis:2016ani}, this scenario was realised in terms of 4d orbifold compactifications with 
chiral matter and spontaneously broken $\mathcal N=1\to 0$ supersymmetry \`a la Scherk-Schwarz, in a simple 
setup involving SO(10) gauge symmetry. There, explicit examples were presented where the one-loop contribution 
is indeed positive, but exponentially suppressed. In recent work \cite{Florakis:2021bws,Florakis:2020rph}, we extended this program by 
introducing string constructions that further break the gauge symmetry to the Pati-Salam level, while preserving the 
super no-scale property $n_B=n_F$. In the same work, the universal features of the one-loop potential were investigated 
and the mismatch in the Bose-Fermi degeneracy of the first massive mode was identified as a parameter controlling the 
positivity of the one-loop potential.
Nevertheless, physical states in these constructions were characterised by a multiplicity of 4, which can be traced 
back to the fact that in each twisted sector, the orbifold shifts exchanged two pairs of fixed points from each of the two
twisted 2-tori, hence, always leaving 4 invariant combinations. As a result, chiral matter states in those 
models were also constrained to occur in multiples of 4, thereby precluding the possibility of constructing three-generation models.

The aim of the present letter is twofold. Namely, we wish to investigate the structure of the one-loop potential in 
more generic setups in which the combined orbifold action shuffles all fixed points in each 2-torus, to precisely remove the extra multiplicities. We will show that the condition $n_B-n_F=0$ considered so far in the string literature \cite{Abel:2015oxa,Kounnas:2016gmz,Aaronson:2016kjm,Ashfaque:2015vta,Kounnas:2017mad,Abel:2017rch,Abel:2017vos,Florakis:2016ani,Florakis:2017ani,Abel:2018zyt,Angelantonj:2019gvi, Abel:2020ldo, Itoyama:2020ifw,Itoyama:2021fwc,Florakis:2021bws}, albeit necessary, is not sufficient to guarantee an exponential suppression 
of the one-loop potential at large radii, but rather needs to be supplemented by additional conditions. Furthermore, we 
will show that a suitable extension of the construction of \cite{Florakis:2021bws} to super no-scale models with 
Pati-Salam gauge symmetry \cite{Pati:1974yy} and 3 generations of chiral matter is possible, and we will illustrate the analysis of the 
extended super no-scale conditions with explicit constructions.


\section{Super No-Scale Conditions}

We consider heterotic orbifold compactifications on $(T^2\times T^2\times T^2)/\Gamma$, where the orbifold group $\Gamma$ is a suitable product of $\mathbb Z_2$ factors. We organize the latter into a product of three orbifold factors $\Gamma=\Gamma_1\times \Gamma_2\times \Gamma_3$.  Here $\Gamma_1=\mathbb Z_2$ realises the Scherk-Schwarz breaking of supersymmetry by acting\footnote{This may be potentially supplemented by additional parity operators associated to the Kac-Moody algebra.} as $(-1)^{F}\delta_2$, where $F$ is the spacetime fermion number and $\delta_2$ is an order-2 shift along a non-trivial cycle of $T^6$ which, for concreteness, is taken along the first 2-torus. Furthermore, we identify $\Gamma_2=\mathbb Z_2\times\mathbb Z_2$ with the non-freely acting orbifold preserving $\mathcal N=1$ supersymmetry and with standard embedding on the gauge degrees of freedom. Finally, $\Gamma_3$ preserves all supercharges of the theory and encodes model-dependent aspects of the construction. In particular, it may contain any additional $\mathbb Z_2$ factors involving translations along the remaining five toroidal directions, twists of the Kac-Moody currents or parity insertions thereof.

The modular  partition function of a generic model in this setup can be cast in the form
\begin{equation}
	Z = \frac{1}{|\Gamma_1|}\sum_{H_1,G_1 \in \Gamma_1} \frac{1}{|\Gamma_2|}\sum_{ {(h_1,h_2)\atop (g_1,g_2)} \in \Gamma_2} \frac{1}{|\Gamma_3|}\sum_{ {(H_2,\ldots)\atop (G_2,\ldots)}\in \Gamma_3} \frac{1}{2}\sum_{a,b=0,1}(-1)^{\Xi}\,\frac{\Lambda\big[ {\textstyle{a\,, \, h_i\atop b\,, \,g_i}}\big]}{\eta^4} \frac{\Gamma_{2,2}^{(1)}\big[ {\textstyle{H_1\,,\,H_2 \atop G_1\,, \, G_2}\big|{h_1\atop g_1}} \big]}{\eta^2\bar\eta^2} \frac{\hat{Z}\big[ {\textstyle{h_i\,, \ldots \atop g_i\,, \ldots}}\big] }{\eta^6\bar\eta^{22}} \,, 
	\label{generalForm}
\end{equation}
where $H_1, (h_1, h_2)$ and $H_2$ label the orbifold sectors of $\Gamma_1, \Gamma_2$ and of a $\mathbb Z_2\subset \Gamma_3$, respectively, while the summation over $G_1, (g_1,g_2)$ and $G_2$ imposes the corresponding projections. Additional parameters labelling the sectors and imposing the projections of any remaining $\mathbb Z_2$ factors in $\Gamma_3$ are not displayed explicitly but denoted as ellipses. $\Lambda/\eta^4$ is the contribution of the worldsheet fermions of the left-moving (RNS) sector of the theory, while $\Gamma_{2,2}^{(1)}/\eta^2\bar\eta^2$ is the (2,2) twisted/shifted lattice encoding the contribution of the Scherk-Schwarz 2-torus with Kähler modulus $T$ and complex structure $U$. In particular, this 2-torus is twisted by the first $\mathbb Z_2 \subset \Gamma_2$ factor, ascribed to $(h_1,g_1)$, which acts as a $\pi$ rotation on both its coordinates. Importantly, the same 2-torus is also shifted by $\delta_2\in\Gamma_1$ and, independently, by $\mathbb Z_2\subset\Gamma_3$, which act as a momentum shift along the first torus direction, and as a winding shift along the second, respectively. Note that, as described in the introduction, the presence of two independent shifts is necessary in order to remove the mod 4 multiplicity of states that would have otherwise been present\footnote{Each of the two twisted 2-tori present in each sector would leave 2 invariant states after modding by $\Gamma_1\times\Gamma_2$, corresponding to the symmetric combinations associated to each of the two pairs of fixed points exchanged under $\delta_2\in\Gamma_1$. Introducing an indepedent shift in $\Gamma_3$ exchanges them and leaves only one invariant state from each 2-torus.}.

The contributions of the remaining two 2-tori, together with those of the right-moving Kac-Moody degrees of freedom, are absorbed into $\hat Z$, modulo their oscillator contributions which we display explicitly. Indeed, we group the latter, together with the non-compact bosons of 4d spacetime, into the factor $1/\eta^{6}\bar\eta^{22}$ involving the Dedekind $\eta(\tau)$-functions, where $\tau=\tau_1+i\tau_2$ is the complex structure of the worldsheet torus.

 Without loss of generality, the phase can be decomposed as $\Xi=a+b+aG_1+bH_1+H_1G_1+\Phi$, with the linear terms encoding spin statistics and imposing the usual GSO projection via the summation over spin structures, whereas the modular invariant term $aG_1+bH_1+H_1G_1$ couples the fermionic R-symmetry charges to the Scherk-Schwarz shift, necessary for the realization of the $\Gamma_1$ action. Possible discrete torsion or additional parity insertions may be implemented by a suitable choice of $\Phi$, which is of course model dependent.

The sum over spin structures may be readily performed using the modified Riemann identity
\begin{equation}
	\frac{1}{2}\sum_{a,b=0,1} (-1)^{a(1+G_1)+b(1+H_1)} \Lambda\big[ {\textstyle{a\,, \, h_i\atop b\,, \,g_i}}\big] = \theta\big[ {\textstyle{1+H_1 \atop 1+G_1} }\big] \, \theta\big[ {\textstyle{1+H_1+h_1 \atop 1+G_1+g_1} }\big]\, \theta\big[ {\textstyle{1+H_1+h_2 \atop 1+G_1+g_2} }\big]\,\theta\big[ {\textstyle{1+H_1-h_1-h_2 \atop 1+G_1-g_1-g_2} }\big]\,,
	\label{RiemannIdent}
\end{equation}
where the r.h.s. is expressed in terms of Jacobi theta constants with characteristics. In the sector $H_1=G_1=0$ where $\Gamma_1$ trivializes, one recovers $\mathcal N=1$ supersymmetry and the partition function vanishes due to the presence of $\theta_1$ in \eqref{RiemannIdent}. Restricting our attention to $(H_1,G_1)\neq(0,0)$, we note a similar vanishing of the partition function also in the sector $(H_1,G_1)=(h_1,g_1)$, which reflects the fact that this sector effectively enjoys the partial spontaneous breaking $\mathcal N=2\to 1$. However, in all remaining sectors $(H_1,G_1)\notin \{(0,0), (h_1,g_1)\}$, the Scherk-Schwarz torus is simultaneously twisted and shifted and the corresponding lattice $\Gamma_{2,2}^{(1)}$ vanishes identically, unless the twist trivializes. Therefore, only the sector $h_1=g_1=0$ where the Scherk-Schwarz torus is untwisted leads to a non-vanishing contribution to the one-loop potential and the partition function may be then cast in the general form
\begin{equation}
	Z = \frac{1}{4}\sum_{  {(H_1,G_1)\atop (H_2,G_2)}\in\mathbb Z_2} \Psi\big[ {\textstyle{H_1\,,\,H_2 \atop G_1\,, \, G_2}}\big](\tau,\bar\tau) \, \Gamma_{2,2}^{(1),\textrm{shift}}\big[ {\textstyle{H_1\,,\,H_2 \atop G_1\,, \, G_2}}\big](T,U;\tau,\bar\tau) \,.
	\label{Zsum}
\end{equation}
Here $\Gamma_{2,2}^{(1),\textrm{shift}}$ is the shifted lattice of modular weight (1,1) associated to the Scherk-Schwarz torus,
\begin{equation}
	\Gamma_{2,2}^{(1),\textrm{shift}}\big[ {\textstyle{H_1\,,\,H_2 \atop G_1\,, \, G_2}}\big](T,U;\tau,\bar\tau) = \sum_{m_i,n_i\in\mathbb Z} e^{i\pi(G_1 m_1+ G_2 n_2)} \, q^{\frac{1}{4}|P_L|^2}\,\bar q^{\frac{1}{4}|P_R|^2}\,,
	\label{ShiftedLattice}
\end{equation}
involving a momentum shift along the first direction and winding shift along the second. Using standard notation, $q=e^{2\pi i\tau}$ denotes the nome, while the complexified lattice momenta 
\begin{equation}
	P_L = \frac{m_2+\frac{H_2}{2}-Um_1+T(n_1+\frac{H_1}{2}+Un_2)}{\sqrt{T_2 U_2}}\ ,\ 
	P_R = \frac{m_2+\frac{H_2}{2}-Um_1+\bar T(n_1+\frac{H_1}{2}+Un_2)}{\sqrt{T_2 U_2}} \,,
\label{ComplexMomenta}
\end{equation}
are expressed in terms of the complex $T,U$ moduli, and the momentum (resp. winding) numbers $m_i$ (resp. $n_i$) run over all integer values.  The particular choice for the embedding of the shifts into the (2,2) lattice in \eqref{ShiftedLattice} is clearly not the most general one. It is, however, sufficient for the purposes of our discussion as well as convenient for rewritting the theories using the framework of the Free Fermionic Formulation (FFF) \cite{Antoniadis:1986rn,Antoniadis:1987wp,Kawai:1986va} later on.

All remaining contributions are assembled into the non-holomorphic modular form $\Psi$ of modular weight $(-2,-2)$. Note that $Z$ itself carries modular weight $(-1,-1)$, since we have not included the contribution $(2\pi\sqrt{\tau_2})^{-2}$ of non-compact bosonic zero modes. Taking the latter into account, we can write the one-loop contribution to the effective potential as the modular integral
\begin{equation}
	V = -\frac{1}{2(2\pi)^4} \int_{\mathcal F} \frac{d^2\tau}{\tau_2^2}\, \frac{Z}{\tau_2} \,,
	\label{Zintegral}
\end{equation}
with $\mathcal F = {\rm SL}(2;\mathbb Z) \backslash \mathcal H$ being the fundamental domain of the modular group and $\mathcal H$ the upper half-plane. The integral is finite, provided one works in non-tachyonic regions of moduli space. This will be the case for our analysis, since we are interested in studying the behavior of the effective potential when the volume of the Scherk-Schwarz 2-torus is sufficiently larger than the string scale. The analytic computation of \eqref{Zintegral}, however, is a daunting task, since the integrand involves a weak non-holomorphic modular form with exponential growth at the cusp, ascribed to the presence of the protograviton term $2/\bar q$, as well as similar terms associated to other unphysical tachyons which are typically present in the Fourier expansion of $Z$. Here, we are interested only in extracting the leading asymptotics in the large volume limit in order to derive the conditions for exponential suppression. 
Following \cite{Angelantonj:2013eja,Angelantonj:2015rxa}, we first decompose the sum \eqref{Zsum} into three independent orbits of $\Gamma_0(2)\backslash {\rm SL}(2;\mathbb Z)$
\begin{equation}
	\tau_2^{-1}Z = \sum_{\alpha\in\{ \rm I,II,III \}} \sum_{\gamma\in \Gamma_0(2)\backslash {\rm SL}(2;\mathbb Z)} Z_{\rm \alpha}\Bigr|_0\gamma \,,
\end{equation}
where $\Gamma_0(2)\subset {\rm SL}(2;\mathbb Z)$ is the Hecke congruence subgroup of the modular group, 
and $|_{0}\gamma$ is the Petersson slash operator acting on the $\Gamma_0(2)$ modular functions $Z_\alpha$, given by
\begin{equation}
\begin{split}
	Z_{\rm I} &= \frac{1}{4} \tau_2^{-1}\Psi\big[ {\textstyle{0\,,\,0 \atop 1\,, \, 0}}\big] \, 
	\Gamma_{2,2}^{(1),\textrm{shift}}\big[ {\textstyle{0\,,\,0 \atop 1\,, \, 0}}\big]\quad\,,\quad
	Z_{\rm II} = \frac{1}{4} \tau_2^{-1}\Psi\big[ {\textstyle{0\,,\,0 \atop 1\,, \, 1}}\big] \, 
	\Gamma_{2,2}^{(1),\textrm{shift}}\big[ {\textstyle{0\,,\,0 \atop 1\,, \, 1}}\big] \,,\\
	Z_{\rm III} &= \frac{1}{2} \tau_2^{-1}\Psi\big[ {\textstyle{0\,,\,1 \atop 1\,, \, +}}\big] \, \Gamma_{2,2}^{(1),\textrm{shift}}\big[ {\textstyle{0\,,\,1 \atop 1\,, \, +}}\big] + \frac{1}{2} \tau_2^{-1}\Psi\big[ {\textstyle{0\,,\,1 \atop 1\,, \, -}}\big] \, \Gamma_{2,2}^{(1),\textrm{shift}}\big[ {\textstyle{0\,,\,1 \atop 1\,, \, -}}\big] \,.\\	
\end{split}\label{ZOrbits}
\end{equation}
In the latter, we defined the combinations $\Psi\big[ {\textstyle{0\,,\,1 \atop 1\,, \, \pm}}\big] =\frac{1}{2}(\Psi\big[ {\textstyle{0\,,\,1 \atop 1\,, \, 0}}\big] \pm \Psi\big[ {\textstyle{0\,,\,1 \atop 1\,, \, 1}}\big] )$ and, similarly for the shifted lattice, $\Gamma_{2,2}^{(1),\textrm{shift}}\big[ {\textstyle{0\,,\,1 \atop 1\,, \, \pm}}\big] = \frac{1}{2}(\Gamma_{2,2}^{(1),\textrm{shift}}\big[ {\textstyle{0\,,\,1 \atop 1\,, \, 0}}\big]\pm\Gamma_{2,2}^{(1),\textrm{shift}}\big[ {\textstyle{0\,,\,1 \atop 1\,, \, 1}}\big])$. Each orbit may now independently be used to partially unfold $\mathcal F$ 
\begin{equation}
	-2(2\pi)^4 V = \sum_{\alpha\in\{ \rm I,II,III \}} \int_{\mathcal F_0(2)} \frac{d^2\tau}{\tau_2^2} Z_{\alpha} \,,
\end{equation}
where each integration is now over the fundamental domain $\mathcal F_0(2) = \Gamma_0(2)\backslash \mathcal H$ of $\Gamma_0(2)$. Each of these integrals may be computed by Poisson-resumming the corresponding lattice
\begin{equation}
	\Gamma_{2,2}^{(1),\textrm{shift}}\big[ {\textstyle{H_1\,,\,H_2 \atop G_1\,, \, G_2}}\big](T,U;\tau,\bar\tau) = \frac{T_2}{\tau_2}\sum_{m_i,n_i\in\mathbb Z}e^{i\pi(0,1)A \binom{G_2}{H_2}+2\pi i\bar T\det(A)-\frac{\pi T_2}{\tau_2 U_2}|(1,U)A\binom{\tau}{1}|^2} \,,
\end{equation}
with
$
	A = \left(\begin{array}{c c}
			n_1+\frac{H_1}{2} & m_1 - \frac{G_1}{2} \\
			n_2 & m_2 \\
			\end{array}\right)\,, 
$
and utilise its decomposition into $\Gamma_0(2)$ orbits in order to unfold $\mathcal F_0(2)$ to the strip or upper half-plane \cite{Dixon:1990pc,Kiritsis:1996xd}. 
Indeed, we first expand
\begin{equation}
	\Psi\big[ {\textstyle{H_1\,,\,H_2 \atop G_1\,, \, G_2}}\big](\tau,\bar\tau) = \sum_{\Delta,\bar\Delta} C\big[ {\textstyle{H_1\,,\,H_2 \atop G_1\,, \, G_2}}\big](\Delta,\bar\Delta)\,q^{\Delta}\,\bar q^{\bar\Delta} \,,
	\label{qqbarExpPsi}
\end{equation}
according to the conformal weights $(\Delta,\bar\Delta)$ of the theory, where $C\big[ {\textstyle{H_1\,,\,H_2 \atop G_1\,, \, G_2}}\big](\Delta,\bar\Delta)$ counts the number of bosonic minus the number of fermionic states with the given weights in each $\big[ {\textstyle{H_1\,,\,H_2 \atop G_1\,, \, G_2}}\big]$ sector. It is then clear that the non-degenerate orbit $\det(A)\neq 0$ always integrates into a modified Bessel function of the second kind and, hence,  leads to exponentially suppressed contributions in $T_2$. Different is the case of the degenerate orbit $\det(A)=0$, which can instead give rise to power law suppression\footnote{The orbit $A=0$ could only arise from the term $H_1=G_1=0$ and, hence, gives a vanishing contribution due to supersymmetry, i.e. $\Psi\big[ {\textstyle{0\,,\,H_2 \atop 0\,, \, G_2}}\big]=0$.}, provided it is integrated against the constant term $C\big[ {\textstyle{H_1\,,\,H_2 \atop G_1\,, \, G_2}}\big](0,0)\equiv C\big[ {\textstyle{H_1\,,\,H_2 \atop G_1\,, \, G_2}}\big]$ in the $q,\bar q$-expansion of eq.~\eqref{qqbarExpPsi}. By inspecting \eqref{ZOrbits}, it is also straightforward to see that we can safely drop the second term in $Z_{\rm III}$, since it projects onto odd $n_2$-windings and, hence, also integrates to exponentially suppressed contributions.

Using the degenerate orbits to unfold $\mathcal F_0(2)$, the problem is reduced to the evaluation of integrals over the half-infinite strip $\Gamma_\infty\backslash\mathcal H$, where $\Gamma_\infty$ is the stabiliser of the cusp at $\infty$. Gathering the various pieces together, we find the power-suppressed contribution to the one-loop effective potential at large volume
\begin{equation}
	V_{\rm eff} = -\frac{63}{2(2\pi)^4\,T_2^2} \left[ \frac{1}{2}\sum_{H_2,G_2\in\mathbb Z_2} (-1)^{H_2} C\big[ {\textstyle{0\,,\,H_2 \atop 1\,, \, G_2}}\big]\, E^\star_\infty(3;U)+ \frac{1}{8}\sum_{G_2\in\mathbb Z_2} C\big[ {\textstyle{0\,,\,1 \atop 1\,, \, G_2}}\big] E^\star_\infty(3;2U) \right] +\ldots\,,
	\label{VeffPower}
\end{equation}
where $E^\star_\infty(s;z)$ is the zero weight, completed, non-holomorphic Eisenstein series\footnote{See, for instance, \cite{Angelantonj:2013eja} and references therein.} of $\Gamma_0(2)$ associated to the cusp at $\infty$
\begin{equation}
	E_\infty^\star(s;z) = \frac{1}{2}\zeta^\star(2s)\sum_{\gamma\in \Gamma_\infty\backslash\Gamma_0(2)} (\Im z)^s\Bigr|_0\gamma = \frac{1}{2}\zeta^\star(2s)\sum_{c,d\in\mathbb Z\atop (2c,d)=1}\frac{(\Im z)^s}{|2c z+d|^{2s}} \,,
	\label{EisenG02}
\end{equation}
 $\zeta^\star(s) = \pi^{-s/2}\,\Gamma(s/2)\zeta(s)$ is the completed Riemann zeta function, and the ellipses in \eqref{VeffPower} denote exponentially suppressed terms. The constraint $(2c,d)=1$ on the r.h.s. of \eqref{EisenG02} restricts the summation to coprime pairs. 
 
 It is now straightforward to read off the super no-scale conditions relevant for our setup. Clearly, \eqref{VeffPower} is suppressed in the volume by a universal $T_2^{-2}$ power, as expected also from dimensional arguments. The $U$-modulus dependence, however, appears as a sum of two independent $\Gamma_0(2)$-Eisenstein series of different arguments. Hence, in order to achieve an exponentially suppressed contribution at large $T_2$, the elimination of the inverse power-law behaviour requires that the coefficients of both Eisenstein series in \eqref{VeffPower} vanish independently. Taking a suitable linear combination of these coefficients, and including also the sector $H_1=G_1=0$ which anyway vanishes by supersymmetry, we can express the two independent conditions compactly as $\Sigma(0)=\Sigma(1)=0$, where
 \begin{equation}\label{super no-scale conditions}
 	\Sigma(H_2) \equiv \frac{1}{4}\sum_{G_1,G_2=0,1} C\big[ {\textstyle{\,0\,\,\,,\,H_2 \atop G_1\,, \, G_2}}\big]  \quad,\ H_2=0,1 \,.
 \end{equation}
It is now straightforward to interpret both conditions as constraints on the massless spectrum of the theory. The first condition $\Sigma(0)=0$, picks the untwisted sector $H_1=H_2=0$ under both freely acting $\mathbb Z_2$'s, projects it onto invariant states by summing over $G_1,G_2$, and finally imposes that the resulting degeneracy of massless bosonic states equal that of massless fermionic ones. It is clear that sectors twisted under either of those $\mathbb Z_2$'s will acquire lattice masses, as can be seen by inspection of the mass formula $M^2(\Gamma_{2,2}^{(1)}) = \frac{1}{2}(|P_L|^2+|P_R|^2)$ and the Hamiltonian representation eqs. \eqref{ShiftedLattice}, \eqref{ComplexMomenta}. As a result, $\Sigma(0)=n_B-n_F=0$ is really a condition on the full massless spectrum of the theory, and is identified with the super no-scale condition known in the string literature. Notice that these massless states are actually accompanied by an infinite tower of light BPS states with $\Delta=\bar\Delta=0$ and masses $M^2(\Gamma_{2,2}^{(1)}) = |m_2-Um_1|^2/T_2 U_2$, which tend to become massless in the infinite volume limit $T_2\to\infty$. Integrating out these states produces the familiar power law suppression of the one-loop potential. As a result, the condition $\Sigma(0)=0$ essentially eliminates their contribution, by exactly balancing the degeneracy of their ground states between bosons and fermions.

The second condition $\Sigma(1)=0$ instead picks the sector $H_1=0$, $H_2=1$, projects it onto invariant states, and imposes a similar cancellation between the degeneracies of bosonic and fermionic BPS states with $\Delta=\bar\Delta=0$. These states are no longer massless, since the $H_2=1$ twist gives a non-trivial contribution to lattice masses $M^2(\Gamma_{2,2}^{(1)}) = |m_2+\frac{1}{2}-Um_1|^2/T_2U_2$. Again, these are states that remain light at large volume and become massless in the asymptotic limit $T_2\to\infty$ where supersymmetry is restored. This second condition is not trivial, and essentially requires that the contribution of this BPS tower of light states also cancels out.

An alternative way to derive these conditions is to realise that, for the purposes of extracting the power law suppression, our unfolding procedure effectively amounts to the strip integral of the constant modes $\Delta=\bar\Delta=0$ in $\Psi$ against the shifted Narain lattice at vanishing windings $n_i=H_1=0$. This is automatically level matched $P_L=P_R$ and the result takes the familiar Schwinger form 
\begin{equation}
	V_{\rm eff} = -\frac{1}{2(2\pi)^4} \sum_{H_2=0,1} \frac{1}{4}\sum_{ G_1, G_2\in \mathbb Z_2} C\big[ {\textstyle{\,0\,\,\,,\,H_2 \atop G_1\,, \, G_2}}\big] \int_{\Gamma_\infty\backslash\mathcal H} \frac{d^2\tau}{\tau_2^2}\,\sum_{P\in\Lambda_{2,2}} (-1)^{m_1}\,e^{-\pi\tau_2 M^2_{\rm BPS}}   +\ldots \,,
\end{equation}
where $M_{\rm BPS}^2 = |m_2+\frac{H_2}{2}-Um_1|^2/T_2 U_2$ is the BPS mass. It is then straightforward to see that the $H_2=0$ and $H_2=1$ sectors integrate into different automorphic functions of the residual T-duality group and the cancellation of the volume suppressed term implies precisely the two independent conditions $\Sigma(0)=\Sigma(1)=0$. The super no-scale condition should hence be stated as the requirement that, in addition to the matching of degeneracies of massless bosons and fermions, also light massive states that become massless in the infinite volume limit must also enjoy a similar matching.


\section{Application in a Class of Pati--Salam Models}

In this section we would like to construct explicit examples of chiral three-generation
heterotic compactifications with spontaneously broken supersymmetry  \`a la Scherk--Schwarz 
and exponentially suppressed one-loop contribution to the effective potential.
To this purpose, we investigate a class of heterotic string models with Pati--Salam $SU(4)\times SU(2)_L\times SU(2)_R$ 
gauge symmetry  \cite{Pati:1974yy, Antoniadis:1988cm, Antoniadis:1990hb, Leontaris:1999ce}.
Following \cite{Florakis:2016ani, Florakis:2021bws} we first employ
the FFF  to define our models and 
analyse 
their spectra (\emph{c.f.} \cite{Gregori:1999ny,Faraggi:2004rq,Faraggi:2006bc,Assel:2009xa}),  and then we map them to orbifold compactifications in order
to implement Scherk--Schwarz supersymmetry breaking and study the moduli dependence of the effective potential.

In the FFF, the class of Pati--Salam models under consideration is defined 
by a basis set of $13$ vectors of boundary conditions for world-sheet fermions
\begin{align}
\begin{split}
v_1=\mathds{1}&=\{\psi^\mu,\
\chi^{1,\dots,6},y^{1,\dots,6},\omega^{1,\dots,6}|\bar{y}^{1,\dots,6},
\bar{\omega}^{1,\dots,6},\bar{\eta}^{1,2,3},
\bar{\psi}^{1,\dots,5},\bar{\phi}^{1,\dots,8}\}\,,\\
v_2=S&=\{\psi^\mu,\chi^{1,\dots,6}\} \quad\,,\quad
v_{2+i}=e_i=\{y^{i},\omega^{i}|\bar{y}^{i},\bar{\omega}^{i}\}\;,i=1,\dots,6\,,\\
v_9=b_1&=\{\chi^{34},\chi^{56},y^{3},y^{4},y^{5},y^{6}|\bar{y}^{3},\bar{y}^{4},\bar{y}^{5},\bar{y}^{6},
\bar{\psi}^{1,\dots,5},\bar{\eta}^1\}\,,\\
v_{10}=b_2&=\{\chi^{12},\chi^{56},y^{1},y^{2},y^{5},y^{6}|\bar{y}^{1},\bar{y}^{2},\bar{y}^{5},\bar{y}^{6},
\bar{\psi}^{1,\dots,5},\bar{\eta}^2\}\,, \\
v_{11}=z_1&=\{\bar{\phi}^{1,\dots,4}\} \quad\,,\quad
v_{12}=z_2=\{\bar{\phi}^{5,\dots,8}\}\quad\,,\quad
v_{13}=\alpha=\{\bar{\psi}^{4,5},\bar{\phi}^{1,2}\}\,,
\end{split}\label{basis}
\end{align}	
and a set of phases, $c_{ij}=c\left[v_i\atop{v_j}\right]=\pm1$, $i\le j=1,\dots,13$, associated with generalised 
GSO (GGSO) projections. In the notation of \eqref{basis}, a world-sheet fermion $f$ transforms as 
$f\to-\exp(i\pi\alpha(f)), \alpha\in\left(-1,+1\right],$ when 
parallel transported along a non-contractible loop of the torus; included fermions are periodic while all the rest are 
antiperiodic. In this language, the basis vectors $\left\{\mathds{1}, S, e_1,\dots,e_6\right\}$ describe an $\mathcal{N}=4$ 
supersymmetric model with $SO(32)$ gauge symmetry. Vectors $b_1, b_2$ are associated with $Z_2\times Z_2$ orbifold  
twists that break supersymmetry to $\mathcal{N}=1$ and the gauge symmetry to $SO(10){\times}U(1)^3{\times}SO(18)$ and vectors
$z_1,z_2$ further reduce gauge symmetry to $SO(10){\times}U(1)^3{\times}SO(8)^2$. The last vector $\alpha$ is utilised 
to  break the $SO(10)$ to the Pati--Salam gauge group.
For generic choices of the GGSO projections this class comprises a huge number of 
$2^{\frac{13(13-1)}{2}+1}$ $\sim 6\times10^{23}$ heterotic string 
models that exhibit 
$ 
{\mathcal G}= SU(4)\times SU(2)_L\times SU(2)_R\times U(1)^3\times SU(2)^4\times SO(8)
$ 
gauge symmetry. Chiral fermion generations transforming as 
$\left(\mathbf{4},\mathbf{2},\mathbf{1}\right)+\left(\overline{\mathbf{4}},\mathbf{1},\mathbf{2}\right)$ under
$SU(4)\times SU(2)_L\times SU(2)_R$ 
arise from 
the sectors $S+b_\alpha 
+ p\,e_i + q\,e_j + r\,e_k + s\,e_l$, where 
$p,q,r,s=\{0,1\}$,  
$a=1,2,3$, with $b_3=b_1+b_2+x$ and $x=\mathds{1}+S+\sum_{i=1}^6 e_i +z_1+z_2$. Pati--Salam gauge symmetry breaking
Higgs scalars  transforming as 
$\left(\mathbf{4},\mathbf{1},\mathbf{2}\right)/\left(\overline{\mathbf{4}},\mathbf{1},\mathbf{2}\right)$ 
 come from the sectors $b_\alpha + p\,e_i + q\,e_j + r\,e_k + s\,e_l$ and Standard Model Higgs fields accommodated in 
 $\left(\mathbf{1},\mathbf{2},\mathbf{2}\right)$ stem from $b_\alpha + x + p\,e_i + q\,e_j + r\,e_k + s\,e_l$. Sectors of the 
form $(S+) b_a+ p\,e_i + q\,e_j + r\,e_k + s\,e_l+\alpha (+z_1)(+x)$ give rise to fractionally charged exotic states transforming in the
$\left(\mathbf{4},\mathbf{1},\mathbf{1}\right)/\left(\overline{\mathbf{4}},\mathbf{1},\mathbf{1}\right)$  and
$\left({\mathbf{1}},\mathbf{1},\mathbf{2}\right)/\left({\mathbf{1}},\mathbf{2},\mathbf{1}\right)$ 
representations of the Pati--Salam gauge group.

We restrict our attention to a subset of these models by imposing a set of conditions on the GGSO phases related to  
a minimal set of requirements. 
These include: (i) Spontaneous breaking of supersymmetry via the Scherk--Schwarz mechanism, translated in the FFF to  
$c[^S_{e_1}]=+1$ supplemented by a set of intricate relationships between the GGSO phases; (ii) Absence of tachyons in the 
physical spectrum of the models; (iii) Presence of at least $3$ complete 
Pati--Salam fermion generations in the massless spectrum; (iv) Existence of at least one massless copy of the Pati--Salam 
and Standard-Model gauge symmetry breaking Higgs bosons; 
(v) No enhancement of the Pati--Salam part of the gauge symmetry;
(vi) Appearance of fractionally charged exotic massless states, common in such compactification
schemes \cite{Wen:1985qj,Athanasiu:1988uj,Schellekens:1989qb,Chang:1996vw,Coriano:2001mg}, as vector-like 
pairs. (vii) The two conditions $\Sigma(H_2)=0$, $H_2=0,1$ as defined in \eqref{super no-scale conditions}, that guarantee the exponential suppression of the one-loop contributions to the cosmological constant in the large volume regime, $T_2\gg1$, along the lines described in  \cite{Florakis:2021bws}.

The detailed formulation of the above criteria in terms of the GGSO coefficients follows the procedure outlined 
in \cite{Florakis:2021bws}, with slight adjustments owed to the modification of the  basis vectors, and will not be 
repeated here. After expressing all these 
constraints in terms of GGSO phases, it turns out that $19$ GGSO coefficients are irrelevant to our analysis
 and can be safely ignored. 
These include the phases $c[^\mathds{1}_\mathds{1}]$ and $c[^\mathds{1}_S]$ which amount to conventions, 
$c[^S_{b_{1}}]$, $c[^S_{b_{2}}]$ and $c[^{b_1}_{b_2}]$ which flip the overall chirality, as well as the phases 
$c[^\mathds{1}_{e_{i}}]$, $i=1,\dots,6$, $c[^\mathds{1}_{b_{1,2}}]$, $c[^\mathds{1}_{z_{1,2}}]$, $c[^{e_{1,2}}_{b_2}]$ 
and $c[^{e_{3,4}}_{b_1}]$ which do not enter into the associated equations. Altogether, we are left with $2^{59}\sim 
5.8\times10^{17}$ in principle distinct non supersymmetric Pati--Salam string models.

To impose the spontaneous supersymmetry breaking and super no-scale conditions, as well as in order to study 
moduli-dependent aspects of the one-loop potential, we shall map the FFF  models to their orbifold 
representation. Although the analysis closely follows that of \cite{Florakis:2016ani}, the difference now is that the 
free worldsheet fermions can no longer be complexified, due to the boundary condition vectors \eqref{basis} explicitly 
involving all six $v_{2+i}$. To this end, we briefly outline the mapping procedure. The orbifold 
partition function corresponding to the free fermion basis \eqref{basis} is of the general form \eqref{generalForm}, 
with the left-moving fermion contribution being
\begin{equation}
	\Lambda\big[ {\textstyle{a\,, \, h_i\atop b\,, \,g_i}}\big] = \vartheta\big[ {\textstyle{a\atop b}}\big] \vartheta\big[ {\textstyle{a+h_1\atop b+g_1}}\big]\vartheta\big[ {\textstyle{a+ h_2\atop b+g_2}}\big] \vartheta\big[ {\textstyle{a-h_1-h_2\atop b-g_1-g_2}}\big] \,,
\end{equation}
and the right-moving one, dressed with the remaining two lattices, being
\begin{equation}
	\begin{split}
	\hat Z\big[ {\textstyle{k\,,\rho \,, h_i\,, H\,, H'\,, H_i\atop \ell\,,\sigma\,, g_i\,, G\,,G'\,,G_i}}\big] =&  \bar\vartheta\big[ {\textstyle{k\atop \ell}}\big]^3\,\bar\vartheta\big[ {\textstyle{k+H'\atop \ell+G'}}\big]\,\bar\vartheta\big[ {\textstyle{k-H'\atop \ell-G'}}\big]\,\bar\vartheta\big[ {\textstyle{k+h_1\atop \ell+g_1}}\big]\,\bar\vartheta\big[ {\textstyle{k+h_2\atop \ell+g_2}}\big]\,\bar\vartheta\big[ {\textstyle{k-h_1-h_2\atop \ell-g_1-g_2}}\big] \\
	\times & \bar\vartheta\big[ {\textstyle{\rho\atop \sigma}}\big]^2 \,\bar\vartheta\big[ {\textstyle{\rho+H'\atop 
	\sigma+G'}}\big] \,\bar\vartheta\big[ {\textstyle{\rho-H'\atop \sigma-G'}}\big]\, \bar\vartheta\big[ 
	{\textstyle{\rho+H\atop \sigma+G}}\big]^4\;
\Gamma_{2,2}^{(2)}\big[ {\textstyle{H_3\,,\,H_4 \atop G_3\,, \, G_4}\big|{h_2\atop g_2}} \big] \, 
	\Gamma_{2,2}^{(3)}\big[ {\textstyle{H_5\,,\,H_6 \atop G_5\,, \, G_6}\big|{h_1+h_2\atop g_1+g_2}} \big]\,,
	\end{split}
\end{equation}
and with the understanding that the additional $\mathbb Z_2$ sums associated to $(k,\ell)$, $(\rho,\sigma)$, $(H,G)$,  
$(H',G')$, $(H_i,G_i)$ are included into the $\Gamma_3$ orbifold factor\footnote{Strictly speaking the parameters 
$(k,\ell), (\rho,\sigma)$ are not ascribed to an orbifold operation, but rather to the fermionic realisation of the 
Kac-Moody lattice of the ${\rm E}_8\times{\rm E}_8$ heterotic string. Nevertheless, we chose to include them into 
$\Gamma_3$ as well in order to keep the general form of \eqref{generalForm}.}. Specifically, the sum over $(k,\ell)$ 
realises the first ${\rm E}_8$ and $(\rho,\sigma)$ the second ${\rm E}_8$ lattice of the ${\rm E}_8\times {\rm E}_8$ 
heterotic string, respectively.  $(h_i,g_i)$ break the first ${\rm E}_8$ factor to ${\rm E}_6\times {\rm U}(1)^3$, 
while $(H,G)$ breaks the hidden (second) ${\rm E}_8$ factor into two $SO(8)$ factors. Finally, $(H',G')$ is responsible 
for further breaking the former to the Pati--Salam gauge symmetry $\mathcal G$. Furthermore, the $(H_i,G_i)$ correspond 
to shifts 
along each of the six directions of internal space with $(H_1,G_1)$ being identified with the Scherk-Schwarz orbifold 
$\Gamma_1$.  Setting all twisted/shifted lattices to their factorised point $T=2U=i$
\begin{equation}
	\Gamma_{2,2}\big[ {\textstyle{H_i\,,\,H_j \atop G_i\,, \, G_j}\big|{h\atop g}} \big](i,\tfrac{i}{2}) = \tfrac{1}{4}\sum_{{\epsilon_i,\,\epsilon_j \atop \zeta_i,\,\zeta_j}\in \mathbb Z_2} \left| \vartheta\big[ {\textstyle{\epsilon_i\atop \zeta_i}}\big]  \vartheta\big[ {\textstyle{\epsilon_i+h\atop \zeta_i+g}}\big] \vartheta\big[ {\textstyle{\epsilon_j\atop \zeta_j}}\big]  \vartheta\big[ {\textstyle{\epsilon_j+h\atop \zeta_j+g}}\big]\right| (-1)^{\epsilon_i G_i+\zeta_i H_i+H_i G_i+\epsilon_j G_j+\zeta_j H_j+H_j G_j} \,,
\end{equation}
the partition function $Z$ can be expressed entirely in terms of genus-one theta constants. Now define $\textbf{a}=(a,a+h_1,a+h_2,a-h_1-h_2,\ldots)^T$ and $\textbf{b}=(b,b+g_1,b+g_2,b-g_1-g_2,\ldots)^T$, to be the vectors of all upper (resp. lower) characteristics of the theta constants entering $Z$. Further denoting by $[\textbf{a}] \equiv \textbf{a}\, {\rm mod}\, 2$ the modulo two operation on $\textbf{a}$, we observe that the set $\{ [\textbf{a}]\}$ of boundary condition vectors for all allowed values of the summation parameters $a,h_1,h_2,\ldots \in \{0,1\}$ is isomorphic to the subgroup of free fermion sets generated by the basis vectors $v_1, \ldots, v_{13}$ in the FFF (and similarly for $\{ [\textbf{b}] \}$). As a result, there exists a 1-1 map between the GGSO phases of the FFF and the corresponding phase $\Xi\big[ {\textstyle{\textbf{a}\atop \textbf{b}}}\big]$ of the orbifold theory. Invariance of the partition function under the generators $T,S$ of ${\rm SL}(2;\mathbb{Z})$, as well as the 2-loop modular transformation $\Omega\to\Omega -\left( {\textstyle{ 0 ~ 1\atop 1~ 0}} \right)$ and requiring factorization into two one-loop amplitudes, the phase is constrained to satisfy 
\begin{equation}
	\begin{split}
		\Xi\big[ {\textstyle{\textbf{a}\atop \textbf{b}+\textbf{a}+\mathds{1}}}\big] = \Xi\big[ {\textstyle{\textbf{a}\atop \textbf{b}}}\big] -1-\tfrac{1}{4}\,\textbf{a}\cdot\textbf{a} \ \,, \ 
		\Xi\big[ {\textstyle{\textbf{b}\atop \textbf{a}}}\big] = \Xi\big[ {\textstyle{\textbf{a}\atop \textbf{b}}}\big]-\tfrac{1}{2}\,\textbf{a}\cdot\textbf{b}\ \,,\ 
		\Xi\big[ {\textstyle{\textbf{a}\atop \textbf{b}+\textbf{b}'}}\big] =\Xi\big[ {\textstyle{\textbf{a}\atop \textbf{b}}}\big]+\Xi\big[ {\textstyle{\textbf{a}\atop \textbf{b}'}}\big]+a\,,
	\end{split}
\end{equation}
where the dot products on the r.h.s. are defined in the Lorentzian sense over the full fermionic charge lattice, and 
the above constraints are required to hold modulo 2. The $a$-shift on the r.h.s. of the third equation above originates 
from the transformation of the worldsheet gravitino determinant at genus two. Taking into account that, modulo 2 in our 
basis, $\frac{1}{4}\textbf{a}\cdot\textbf{a}=a^2-H^2+{H'}^2$ and $\frac{1}{2}\textbf{a}\cdot\textbf{b}=0$, the above 
constraints can be uniquely solved by setting $\Xi\big[ {\textstyle{\textbf{a}\atop 
\textbf{b}}}\big]=a+b+HG+H'G'+\Phi\big[ {\textstyle{\textbf{a}\atop \textbf{b}}}\big]$, where $\Phi\big[ 
{\textstyle{\textbf{a}\atop \textbf{b}}}\big]=\textbf{X}^T \textbf{M}\textbf{Y}$ is constructed as a modular invariant 
bilinear in the parameters $\textbf{X}^T=(a,k,\rho,\epsilon_i,h_i,H,H')$ and 
$\textbf{Y}^T=(b,\ell,\sigma,\zeta_i,g_i,G,G')$. Here, the $13\times 13$ matrix $\textbf{M}$ with $\mathbb Z_2$ entries 
is symmetric and constrained to satisfy the 12 conditions\footnote{In general, for $n$ basis vectors there are 
$\frac{1}{2}n(n-1)$ conditions from the symmetry of $\textbf{M}$ but only $(n-1)$ conditions from the T-transformation, 
since $\sum_i \sum_{j(\neq i)}M_{ij}=0$ (mod 2) trivially. The number of independent choices is then 
$2^{\frac{n(n-1)}{2}+1}$, in exact agreement with the fermionic formulation.} $\sum_{j(\neq i)}M_{ij}=0$ (mod 2). To 
uniquely determine $\textbf{M}$, it is sufficient to evaluate its independent elements for choices $\textbf{X}$, 
$\textbf{Y}$ corresponding to the basis vectors of the fermionic construction and identify it with the corresponding 
GGSO phases. Some care is required in matching the two representations of $Z$, since the fermionic formulation utilises 
the $\mathbb Z_2$ vectors $\{ [\textbf{a}] \}$, whereas the characteristics of the theta constants appearing in 
\eqref{generalForm} are written in terms of the set of integer vectors $\{\textbf{a}\}$. To account for the extra 
phases arising from periodicities of the theta constants,  the additional phase factor $\Lambda\big[ 
{\textstyle{\textbf{a}\atop \textbf{b}}}\big]=\exp [\frac{i\pi}{2}(\textbf{a}-[\textbf{a}])\cdot \textbf{b}]$ needs to 
be inserted, when comparing with the FFF setup. Note that, depending on the problem at hand, such as for 
studies of the one-loop potential, other representations of $\Lambda$ may also be available, obtained after exploiting 
theta identities.

\begin{figure}[h!]
	\centering
	\includegraphics[scale=0.8]{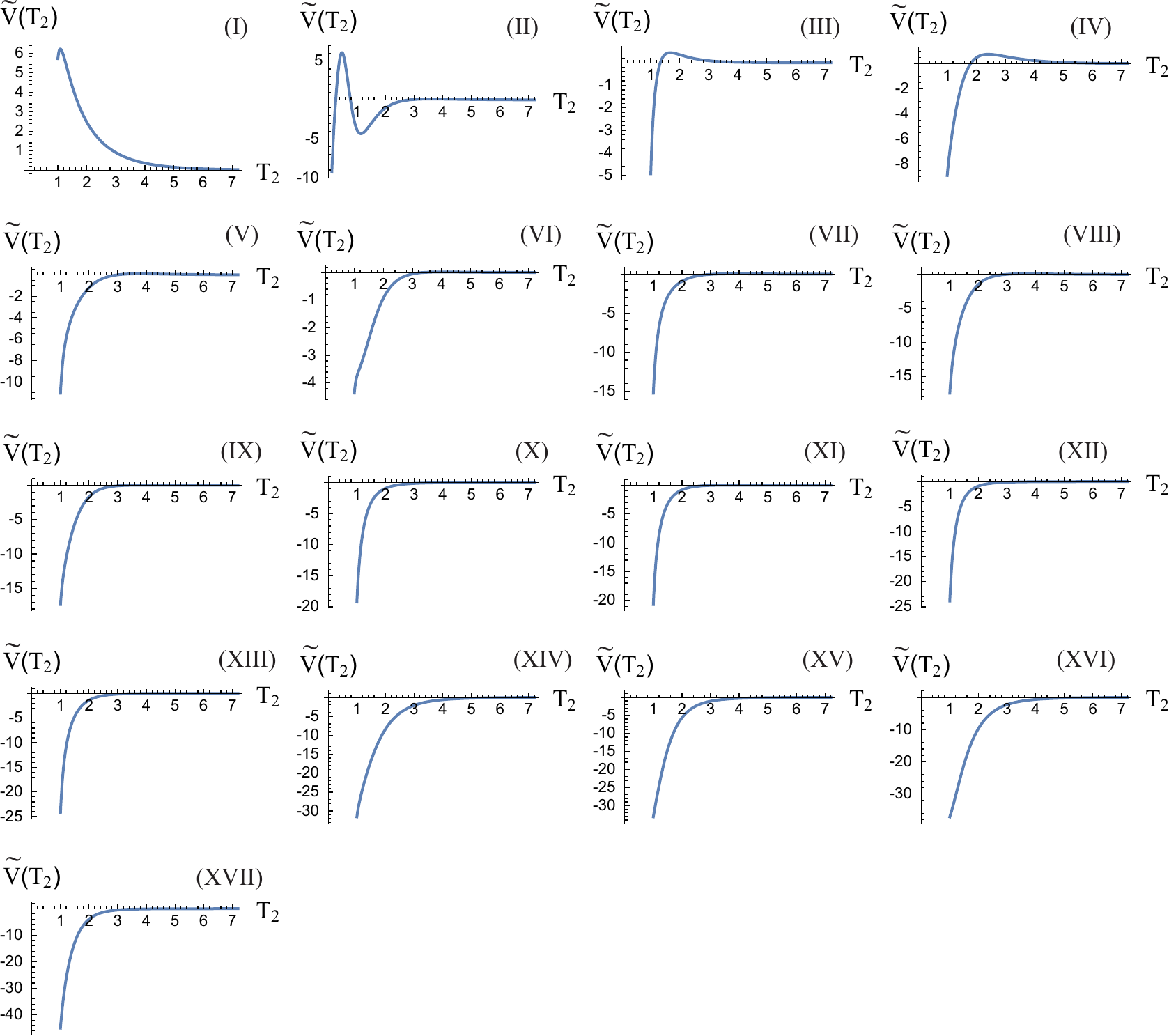}
	\caption{The rescaled one-loop effective potential $\tilde{V}(T_2)=2(2\pi)^4V(T_2)$ for each of the $17$ classes of 
	models satisfying all conditions (i)-(vii).}\label{potentials}
\end{figure}

We now proceed with a detailed investigation of the parameter space of the models utilising 
a computer assisted two-stage scan procedure \cite{Assel:2010wj,Faraggi:2017cnh,Faraggi:2019qoq,Faraggi:2020wld,Florakis:2021bws}. 
In this approach, we first perform a (random) scan of the configurations generated 
by the first 12 basis vectors in \eqref{basis} and identify  $SO(10)$ parent configurations compatible with 
our search criteria.  For each such configuration we consider all possible
offspring Pati--Salam models generated after the introduction of the vector $v_{13}=\alpha$ and
the related GGSO projection phases and check their compatibility with the aforementioned criteria.
In practice, this method allows us to effectively scan a big sample of $8.1\times10^{12}$ models (almost one model in 
$10^4$) of the full parameter space in about 10 days on a DELL PowerEdge R630 workstation with 32 GB of memory.
It turns out that $8.8\times10^6$ models fulfil criteria (i)-(vi). Out of these, about $0.1\%$ meet
the first super no-scale constraint (vii), $\Sigma(0)=0$, while around $21\%$ meet the second super no-scale constraint (vii), $\Sigma(1)=0$.
Altogether,  we identify $174$ Pati--Salam models that comply with all requirements (i)-(vii).
As far as the one-loop effective potential is concerned, the models fall into $17$ distinct classes based on the 
analysis of their partition functions 
\eqref{generalForm}.
An overview of their potentials is shown in Figure \ref{potentials} where we depict the
one-loop effective potential $\tilde{V}(T_2)=2(2\pi)^4V(T_2)$ as a function of $T_2$ modulus of the 
Scherk--Schwarz torus for each class (I)-(XVII) while keeping all remaining moduli to their fermionic point values.
All potentials are finite in the plotted region and, with the exception of class (II), they develop tachyons
for $T_2<1$. Note that the additional lattice shifts employed here spoil the simple T-duality symmetry $T_2\to 1/T_2$ of \cite{Florakis:2016ani,Florakis:2021bws}.

An exemplary model that satisfies all criteria (i)-(vii), henceforth referred to as Model A, is defined by the GGSO 
matrix of Eq. \eqref{ModelAB} (upper signs, where relevant)
\footnote{The lower signs in \eqref{ModelAB} when present, correspond to Model B discussed below.}
\begin{equation}
\begin{small}
\label{ModelAB}
c[^{v_i}_{v_j}]=
\left(
\begin{array}{ccccccccccccc}
+  & +  & +  & +  & +  & +  & +  & +  & +  & +  & +  & +  & -  \\
+  & +  & +  & -  & -  & -  &  \pm   & -  & +  & +  &  \mp   & +  &  \pm  \\
+  & +  & -  & +  & +  & +  &  \mp   & +  & -  & +  &  \pm   & -  &  \mp   \\
+  & -  & +  & -  &  \mp   & -  &  \pm   &  \mp   &  \mp   & +  & -  & +  &  \mp   \\
+  & -  & +  &  \mp   & -  & +  &  \mp   &  \pm   & +  &  \pm   &  \pm   & +  & +  \\
+  & -  & +  & -  & +  & -  & -  &  \pm   & +  &  \pm   &  \pm   & +  & \pm   \\
+  &  \pm   &  \mp   &  \pm   &  \mp   & -  & -  & -  & -  & +  &  \mp   &  \mp   &  \mp   
\\
+  & -  & +  &  \mp   &  \pm   &  \pm   & -  & -  & +  &  \pm   &  \mp   &  \mp   &  \mp   
\\
+  & -  & -  &  \mp   & +  & +  & -  & +  & +  & +  & -  &  \pm   &  \mp   \\
+  & -  & +  & +  &  \pm   &  \pm   & +  &  \pm   & +  & +  & -  & -  &  \pm   \\
+  &  \mp   &  \pm   & -  &  \pm   &  \pm   &  \mp   &  \mp   & -  & -  & +  &  \pm   & -  
\\
+  & +  & -  & +  & +  & +  &  \mp   &  \mp   &  \pm   & -  &  \pm   & +  &  \mp   \\
-  &  \pm   &  \mp   &  \mp   & +  &  \pm   &  \mp   &  \mp   &  \pm   &  \mp   & +  &  
\mp   & -  \\
\end{array}
\right) \,.
\end{small}
\end{equation}
This model includes three fermion generations arising in the sectors: 
$S+b_1+e_6$, $S+b_1+e_3+e_4+e_6$, $S+b_2+e_1+e_6$, $S+b_2+e_2+e_5+e_6$, $S+b_3+e_1$, $S+b_3+e_1+e_3+e_4$, in addition 
to heavy and light Higgs scalars from sectors $b_2+e_6$ and $b_1+e_6+x$, $b_1+e_3+e_4+e_6+x$ respectively. It does not 
exhibit gauge symmetry enhancements, and the exotic fermions in its massless spectrum are organised into vector-like 
pairs. Furthermore, SUSY breaking is consistent with the stringy realisation of the Scherk--Schwarz mechanism outlined 
above. As advertised this model satisfies both super no-scale conditions (vii) and thus exhibits an 
exponential suppression of the one-loop effective potential when $T_2\gg1$, driving the theory towards a low SUSY 
breaking scale and a sufficiently small cosmological constant. The potential falls into class (I) in Figure \ref{potentials} and  
exhibits a global maximum near the 
fermionic point, while the region $T_2<1$ contains tachyons.
The partition function at generic points of the moduli space can be obtained from \eqref{generalForm}, using the 
orbifold phase:
\begin{equation}
\begin{aligned}
\Phi\big[ {\textstyle{\textbf{a}\atop \textbf{b}}}\big]&=\;ab+a(G_1+g_1)+b(H_1+h_1)+k\ell+k(G_1+g_1)+\ell(H_1+h_1)\\
&+\rho\sigma+\rho(G+G'+G_1+G_3)+\sigma(H+H'+H_1+H_3)\\
&+HG+H(G'+G_1+G_3+G_6+g_1+g_2)+G(H'+H_1+H_3+H_6+h_1+h_2)\\
&+H'G'+H'(G_2+G_3+G_5+G_6+g_2)+G'(H_2+H_3+H_5+H_6+h_2)\\
&+H_1G_1+H_1(G_2+G_3+G_5+g_1+g_2)+G_1(H_2+H_3+H_5+h_1+h_2)+H_2(G_5+g_2)\\
&+G_2(H_5+h_2)+H_3G_3+H_3(G_4+G_5+G_6)+G_3(H_4+H_5+H_6)+H_5g_2+G_5h_2\,.
\end{aligned}
\end{equation}

\begin{figure}[h!]
	\includegraphics[width=15cm]{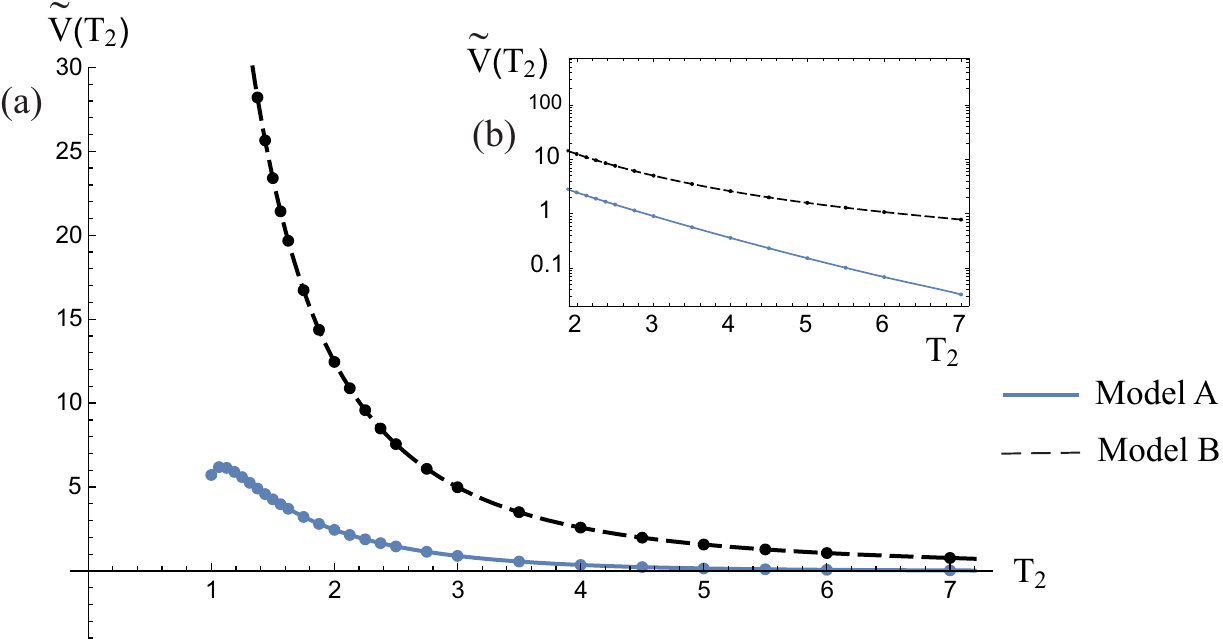}
	\caption{Comparison between the rescaled one-loop effective potentials $\tilde{V}(T_2)=2(2\pi)^4V(T_2)$ of 
	Models A and B in linear (a) and detail in semi-logarithmic (b) scale, showing  the exponential suppression present 
	in Model A  as opposed to Model B.}
\label{figure2}
\end{figure}

In order to highlight the necessity of both super no-scale conditions \eqref{super no-scale conditions} in the 
construction of models with an exponentially suppressed cosmological constant, we proceed to compare its potential 
with a model which satisfies only the usual $\Sigma(0)=0$ super no-scale condition. 
To this end, consider Model B defined by the GGSO 
matrix of Eq. \eqref{ModelAB} (lower elements).
This Pati--Salam model exhibits three fermion generations arising from the sectors $S+b_1$, $S+b_1+e_6$, 
$S+b_3+e_1+e_2+e_3+e_4$, $S+b_1+e_3+e_4+e_5$, $S+b_1+e_3+e_4+e_5+e_6$ and $S+b_2+e_6$, as well as the necessary scalars 
needed for the spontaneous breaking of the Pati--Salam and Standard Model gauge symmetries in the 
sectors $b_2+e_1+e_6$ and $b_2+e_1+e_2+e_5+e_6+x$, $b_3+e_2+e_3+x$ respectively. Again, the exotic fermion content is 
organised into vector-like pairs, while no additional vector bosons leading to gauge symmetry enhancement are present. 
Supersymmetry breaking can be traced to the Scherk--Schwarz mechanism implemented as an orbifold shift along the first 
compactified direction. In addition to satisfying criteria (i)-(vi), Model B satisfies the first of the super 
no-scale condition  $\Sigma(0)=0$ , but fails to satisfy the new condition  $\Sigma(1)=0$. As a result, the one-loop 
effective 
potential as a function of the 
$T_2$ modulus of the Scherk--Schwarz torus is positive semi-definite, but does not exhibit the desired exponential 
suppression when $T_2\gg1$. A comparison of the potentials of the two models is shown in Figure \ref{figure2}. 

\FloatBarrier
\FloatBarrier
\FloatBarrier
\FloatBarrier

\newpage 
	
\section*{Acknowledgments}
	
The research of K.V. is co-financed by Greece and the European Union (European Social Fund - ESF) through the Operational Programme ``Human Resources Development, Education and Lifelong Learning" in the context of the project ``Strengthening Human Resources Research Potential via Doctorate Research" (MIS-5000432), implemented by the State Scholarships Foundation (IKY).

\bibliographystyle{JHEP1}

\providecommand{\href}[2]{#2}\begingroup\raggedright\endgroup

\end{document}